\documentclass[11pt]{article}

\usepackage[english]{babel}
 \usepackage{cite,wrapfig}
 \usepackage{amsfonts}
 \usepackage{graphicx, epstopdf}
 \usepackage[useregional]{datetime2}
 \usepackage{graphics}
 \usepackage{amsthm}
 \usepackage{amssymb}
 \usepackage{amsmath}
 \usepackage{caption}
 \usepackage{bbm}
 \usepackage[section]{placeins}
\usepackage{algorithm}
 \usepackage{algorithmic}
 \usepackage{multirow}
 \usepackage{subfig}
 \usepackage[a-1b]{pdfx}
\usepackage{hyperref}
 \usepackage{xcolor}
\usepackage{booktabs}
 \usepackage{array}
\usepackage{verbatim}
\usepackage{float}
\usepackage{placeins}
\usepackage{stackengine}
\usepackage[tablename=TABLE]{caption}
\usepackage[figurename=FIG.]{caption}

\def\1{\mathbbm{1}}

\def\0{\mathbf{0}}

\newtheorem{lemma}{Lemma}[section]

\theoremstyle{definition}
\newtheorem{definition}{Definition}[section]

\theoremstyle{definition}

\theoremstyle{remark}

\title{Genomics Data Analysis via Spectral Shape and Topology}
\author{Erik J. Am\'{e}zquita\thanks{Computational Mathematics, Science, and Engineering, Michigan State University}
\and Farzana Nasrin  \thanks{Department of Mathematics, University of Hawaii at Manoa}
\and Kathleen M. Storey \thanks{Department of Mathematics, Lafayette College}
\and Masato Yoshizawa\thanks{School of Life Sciences, University of Hawaii at Manoa}
}

\date{}
\begin{document}
\maketitle
\providecommand{\keywords}[1]{\textbf{\textit{Keywords}} #1}

\begin{abstract}
 Mapper, a topological algorithm, is frequently used as an exploratory tool to build a graphical representation of data. 
 This representation can help to gain a better understanding of the intrinsic shape of high-dimensional genomic data and to retain information that may be lost using standard dimension-reduction algorithms.  
	     We propose a novel workflow to process and analyze RNA-seq data from tumor and healthy subjects integrating Mapper and differential gene expression. Precisely, we show that   a
	  Gaussian mixture approximation method can be used to produce graphical structures that successfully separate tumor and healthy subjects, and produce two subgroups of tumor subjects. A further analysis using \emph{DESeq2}, a popular tool for the detection of differentially expressed genes, shows that these two subgroups of tumor cells bear two distinct gene regulations, suggesting two discrete paths for forming the lung cancer, which could not be highlighted by other popular
clustering methods, including t-SNE. Although Mapper shows promise in analyzing high-dimensional data,  building tools to statistically analyze Mapper graphical structures is limited in the existing literature. In this paper, we develop a scoring method using \emph{heat kernel signatures} that provides an empirical setting
      for statistical inferences such as hypothesis testing, sensitivity analysis, and correlation analysis.     
	     
\keywords{Differential gene expressions, Gaussian mixture, heat kernel signature, Mapper,  RNA-sequencing data.}
\end{abstract}

\section{Introduction}
Topological data analysis (TDA) is a mathematical approach that has yielded promising results to unravel the underlying structure of diverse data sets in biology. At the molecular level, TDA has been used to understand the structure of proteins \cite{Kovacev2016}, protein-ligand binding affinities \cite{Cang2018a, Cang2018b}, and viral reassortment \cite{Chan2013}. M.~Nicolau and coauthors \cite{Nicolau2011}, used TDA to identify a previously unreported group of breast cancer tumors with a unique molecular profile and excellent prognosis. They employed a topological algorithm, known as Mapper, to build a graphical representation of the data that reduces the dimensionality of the data while still preserving its local structure \cite{Singh2007}. J.~Arsuaga et al.~developed a homology-based classification method for genomic hybridization arrays and gene expression in \cite{DeWoskin2010, Arsuaga2012}, and showed that it could distinguish most breast cancer subtypes. They extended this topological method and discovered new DNA copy number aberrations within specific subtypes of breast cancer \cite{Arsuaga2015}. A recent study by R.~Jeitziner et al. \cite{Jeitziner2019} presented a new two-tiered version of the Mapper algorithm, which is particularly useful for small genomic sample sizes. Topological approaches to RNA-sequencing (RNA-seq) data have also been applied to study the \textit{in vitro} differentiation of murine embryonic stem cells into neurons in \cite{Rizvi2017}, suggesting the translatability of Mapper-based topological analysis to many biological contexts.

There remains a massive quantity of high-throughput data unexplored by TDA, so we have begun our investigation by applying a Mapper-based classification approach to genomic data sequenced from lung adenocarcinoma samples. We choose to focus on lung carcinoma, since it is frequently accompanied by a large number of genetic mutations. The set of RNA-seq cancer data is provided in the Cancer Genome Atlas (TCGA), which is an effort run by the National Cancer Institute and the National Human Genome Research Institute to provide a vast amount of open access data from over 11,000 cancer patients \cite{TCGA}. The  lung adenocarcinoma data was initially collected in two studies from 2014 and 2016 \cite{lung2016, lung2014}. We compare the data sequenced from lung tumor tissue with data sequenced from healthy lung tissue. The healthy tissue data is provided in the Genotype Tissue Expression (GTEx) project, which has cataloged over 9000 healthy tissue samples \cite{Gtex2013,Gtex2015}.

We propose a novel workflow to describe and analyze bulk tumor cell RNA-seq data using the Mapper algorithm, in order to detect robust patterns in the genetic data. This workflow involves first describing the data using a Gaussian mixture approximation and corresponding scores, which facilitates comparison between subjects.  The Gaussian fitting-based analysis of RNA-seq data was proposed in \cite{Hart2013}. However, the method relies on a predefined cutoff point generated by ChiP-seq data, which is not available for our analysis. To that end, we implement a Gaussian mixture model to fit the data and use Mapper to construct a graphical representation of the Gaussian mixture scores. 
We apply this workflow to the lung data set described above, and we identify two distinct groups of tumor subjects that are consistently separated graphically by healthy subjects. 
In order to determine which subjects are placed consistently in the graph as Mapper parameters vary, we assign an index to each subject, which we call the \emph{position index (PI)}.

In this work, we develop a further analysis of Mapper graphical representation obtained from the RNA-seq data guided by a pipeline of gene expression comparisons, using \emph{DESeq2}, followed by gene pathway prediction with Gene Ontogeny (GO) analyses with Enrichr. DESeq2 is built on  negative binomial generalized linear models and is useful for the detection of  differentially expressed genes \cite{love2014}. We use DESeq2 to perform analysis on position index (PI)-specific RNA-seq count data in a pairwise manner between
each of the three subgroups generated by Mapper - one mostly consisting of  healthy subjects vs. two others composed primarily of tumor subjects.  The analysis shows that the majority of the $p$-values are less than the significance level of
$0.05$, thus indicating that the two tumor subgroups have significantly large numbers of dysregulated genes.
We then use Enrichr to identify the enriched GO terms and molecular pathways among these differentially expressed genes \cite{Xie2021,Kuleshov2016,Chen2013}. This analysis predicts the biological processes that these dysregulated genes belong to, such as inflammatory reactions and muscle functions, pointing to possible impaired biological processes in cancer patients, thereby informing targeted therapeutic approaches.

Finally, we develop a scoring method to provide statistical inferences, in particular, sensitivity analysis under different parameter choices. We denote these scores as graphical subject scores (GSS).  Although we focus on sensitivity analysis in this paper, the GSS method is general enough to be employed for other statistical inferences such as correlation analysis and hypothesis testing. The scoring method is built on \emph{heat kernel signatures} (HKS) which can be thought of as spectral node signatures of graphs \cite{Jian2009}.  HKS  belongs to a family of spectral node signatures called Laplacian family signatures (LFS). LFS is parameterized by a construction filter that relies on graph Laplacian \cite{Hu2014}. LFS has drawn recent attention for graph matching and graph classification problems \cite{Chowdhury2021, Royer2021, carriere2020, Yim2021}. In particular, the integration of spectral shape analysis and topological tools has been introduced in recent work \cite{Royer2021, carriere2020}. The combination of topological structures and HKS proves to be useful, as the former can capture the global topological properties, whereas the latter relies on the  values generated by the spectral signatures to characterize the graphical structure, but HKS cannot be used to extract the topological information on its own. We consider the Mapper graphical structure as a \emph{doubly weighted graph},  and the weighted graph Laplacian is based on the definition in \cite{Xu2020}. The HKS assigns scores to each node  and consequently to each subject in the Mapper graph. The main contributions of this work are: 
\begin{enumerate}
    \item A novel framework to integrate the Mapper graphical
structures and traditional differential gene expression analysis to provide insight regarding distinctive genetic markers. 
\item A scoring method based on spectral shape analysis to develop an empirical setting for statistical inferences of Mapper graphical structures.  
\end{enumerate}

 This paper is organized as follows. 
  Section \ref{sec:prelm} provides necessary background and definitions for building the methodology of the paper. 
  In Section
 \ref{sec:main}, we describe the key methodologies. 
 Detailed demonstrations of the data set and results using the methodologies 
  are presented in Section \ref{sec:results}.   
 Finally, we end with a discussion in Section \ref{sec:discussion}. Additional details describing the genetic analysis results are provided in supplementary materials.

\section{Preliminaries} \label{sec:prelm}

We begin by discussing the necessary background for generating Mapper graphical structures and GSSs. In Subsection \ref{subsec:gm scores}, we  review the Gaussian mixture scores which are used for constructing Mapper graphs. 
Pertinent definitions, a lemma, and some basic facts about HKS are discussed in Subsection \ref{subsec:prelm hks}.

\subsection{Gaussian mixture scores \label{subsec:gm scores}} 
Let $x$ be a one-dimensional vector generated from a Gaussian mixture model (GMM) of $N$ components obtained by fitting the  distribution of FPKM values. Then the probability density function for the GMM is given by  
\begin{equation} \label{eqn:gauss_mix_model}
    f(x) = \sum_{j = 1}^{N}c_{j}\mathcal{N}(x;\mu_{j},\sigma_{j}), 
    \vspace{-0.1in}
    \end{equation}
    where $N$ is the number of mixture components, and $c_j, \mu_j$, and $\sigma_j$ are the weights, means, and standard deviations, respectively. In order to standardize the Gaussian mixture \cite{li2018}, we implement the notion of \emph{membership weight} in Bayesian probability. Precisely, if $P(\pi = j) = c_k$ for a latent variable $\pi \in \{1, \cdots, N\}$ assign the class of $x$, we have $x| \pi  \sim \mathcal{N}(\mu_{\pi}, \sigma_{\pi})$.  An equivalent definition may arise by using the indicator function $\mathbbm{1}$ for the latent variable. Let $s_j = \mathbbm{1}_{ \pi = j}$, then for $s \sim \text{Multinominal}(1;c_1, \cdots, c_N) $,  $x| s \sim \mathcal{N}(\sum_{j=1}^N s_j\mu_{j}, \sum_{j=1}^N s_j\sigma_{j})$. The membership weight is then defined by 
    
    \begin{equation}
        \omega_j = \frac{c_j y_j}{\sum_{j=1}^N c_j y_j},
    \end{equation}
where $y_j(\cdot) = y_j(\cdot|\mu_{j},\sigma_{j}) $. In order to estimate the latent variable, the authors of \cite{li2018} discuss two types of assignment - hard and soft assignment,  and by relying on these assignments they provide four standardized scores. We adopt their notations to define these scores. Let $Z \sim \mathcal{N}(0,1)$, $\Delta = \sqrt(\sigma)(\mu_1-\mu_2)$, $\tau = \sqrt{\frac{\sigma_1}{\sigma_2}}$, and $\Tilde{s}_j$ be the estimate of $s_j$ then
\begin{align}
    T_0 & = Z \mathbbm{1}_{\Tilde{s}_1 = s_1}+(\tau^{-1}Z-\Delta_1)\mathbbm{1}_{\Tilde{s}_1 > s_1}+ (\tau Z + \tau \Delta_1)\mathbbm{1}_{\Tilde{s}_1 < s_1} \label{eqn:T0}\\
     T_1 & = \big(\sum_{j=1}^N \Tilde{s}_j \sqrt{\sigma_1}\big) \big(x-\sum_{j=1}^N \Tilde{s}_j \mu_j\big) \label{eqn:T1}\\
     T_2 & = \big(\sum_{j=1}^N \Tilde{s}_j {\sigma_1}\big)^{-1/2} \big(x-\sum_{j=1}^N \Tilde{s}_j \mu_j\big) \label{eqn:T2}\\
     T_3 & = (\sum_{j=1}^N \Tilde{s}_j[\sigma_j+(\mu_j-\sum_{j=1}^k\Tilde{s}_j \mu_j)^2])^{-1/2}(x - \sum_{j=1}^N \Tilde{s}_j \mu_j) \label{eqn:T3}
\end{align}

\subsection{Heat Kernel Signatures (HKS) \label{subsec:prelm hks}}
\begin{definition} \cite{Xu2020} (Doubly-weighted graph).
A connected undirected graph  $G = (V, M, W)$ is a doubly weighted graph, where $V = \{1, \cdots, n\}$ is the vertex set, $M = diag \{m_1, \cdots, m_n\}$ is the diagonal matrix for weights of vertices, and $W =  \{W_{ij}\}$ is the matrix for weights of edges. If $D = \{d_1, \cdots, d_n\}$ is the degree matrix of the graph $G$, then $d_i = \sum_j W_{ij}$. 
\end{definition}

\begin{definition} \cite{Xu2020} (Weighted graph Laplacian).
Suppose $G = (V, M, W)$ is  a doubly weighted graph. Let $\mathcal{G}$ be the linear space of all functions $f: V \rightarrow \mathbb{R}$. The gradient of $f$ is defined as a vector $\nabla f : = (f(y)-f(x))\sqrt{\frac{W_{xy}}{m_x}}$ for all $y \in V$, and the weighted Laplacian $\Delta$ is an operator in $\mathcal{G}$ defined as  $\Delta f : = \sum_{y \in V} (f(y)-f(x))\sqrt{\frac{W_{xy}}{m_x}}$. The integral of $f$ is defined as $\int f :=  \sum_{x\in V} f(x) m_x$, and $\mathcal{G}$ is equipped with the inner product $<f,g> = \int fg$ for all $f,g \in \mathcal{G}$. 
\end{definition}

\begin{lemma} \cite{Xu2020}
$\Delta$ is equivalent to the weighted Laplacian matrix $L_{M,W} = M^{-1/2}(D-W)M^{-1/2} \in \mathbb{R}^{n \times n}$. Notice that when $M = I$ or $M = D$, a weighted Laplacian becomes an unnormalized
Laplacian or normalized Laplacian. 
\end{lemma}

The Laplacian matrix is symmetric and positive semi-definite. 
However, this is true in general, as in \cite{Xu2020} the authors provide the weighted spectral algorithm and  show how to compute the eigendecomposition for $L_{M,W}$. They denote the eigenpairs by $\{\lambda_k, \phi_k\}_{k=1}^{|V|}$. 

\begin{definition} \cite{Hu2014} (Laplacian Family Signature (LFS).)
Suppose $h(\cdot; \cdot): \mathbb{R}^2 \rightarrow \mathbb{R}$ is the construction filter function for a family of signatures. Then the LFS of a node $i \in V$ is a one-parameter family of structural node descriptors,

\begin{equation} \label{eq:hks}
s_i(t) = \sum_k h(t;\lambda_k) \phi_k(i)^2.
\end{equation}

\end{definition}

As the signature of a given node $i \in V$ is a function of the parameter analogous to time in the well-known heat diffusion process, two nodes $i$ and $j$  can be compared using any kind
of distance or norm between the functions $s_i(\cdot)$ and $s_j(\cdot)$.
A physical interpretation of the node signature is that it captures the amount of heat left at the node at various times $t$, assuming  initially $t = 0$. To obtain HKS, we select $ h(t;\lambda_k) = \exp(-t\lambda_k)$. 

\section{Methodology \label{sec:main}}

\subsection{Gaussian mixture approximation of FPKM data \label{sec:GM approx}} 

When analyzing RNA-seq data, it is routine to differentiate high and low expression genes, or detect proteins from several genes expressed below an appropriate threshold \cite{Hebenstreit2011, Hart2013, Mortazavi2008}. By using visual interpretation and curve fitting to assess the quality of fitting,  Hebenstreit and collaborators \cite{Hebenstreit2011} showed that RNA-seq follows a bimodal distribution of high and
low expression genes. In \cite{Mortazavi2008}, the authors  calculated an automatic measure of the presence and prevalence of
transcripts from known and previously unknown genes. FPKM (Fragments Per Kilobase Million) computation is gaining attention as an expression for RNA-seq data, as this measure is normalized with respect to gene length and allows us to identify the relative gene expression more intuitively. 

An alternative to the use of FPKM values, in particular the Z-scores of FPKM  values,  was proposed in \cite{Hart2013}. However, they rely on a predefined threshold as a cutoff point and another set of data known as ChiP-seq data to produce the ratio of active to repressed promoters. In order to  find the  Z-scores of FPKM, they fit a Gaussian curve to 
each gene expression curve centered at the half-point of the $\log_2(\text{FPKM})$ values. The fitting of a symmetric distribution to an asymmetric curve is justified by the cutoff point, as anything to the left of that point was ignored. In our study, the  ChiP-seq data is not available, so there is no reasonable cutoff point to consider. However, the Z-score approach proves to be useful if the goal is to visualize the gene expression curve centering the highest expressed genes for each individual. In order to capture the full distribution of gene expression levels, we choose to implement the Gaussian mixture model to fit the $\log_2(\text{FPKM})$ values. Next we present an example as a justification for applying this model.

\begin{figure}[htbp]
				\centering
			
				\includegraphics[width=6in,height=2in]{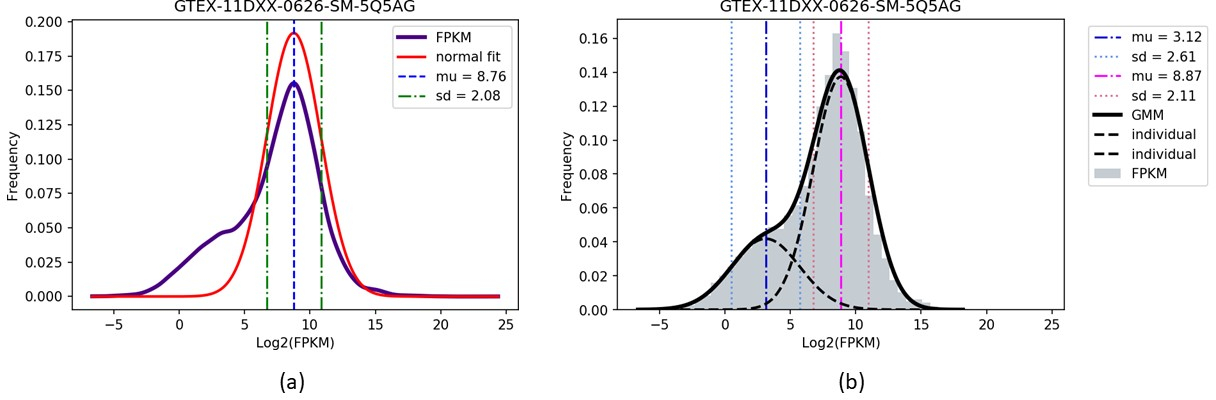}
				\caption{An example of (a) the fitted Gaussian curve  and  (b) the fitted Gaussian mixture curve of two components to the distribution of $\log_2(\text{FPKM})$ values for a sample subject from the GTEx data set. \label{fig:Gauss_mix_exp}}
			\end{figure}

\subsubsection{Motivating Example \label{motivating_example}} 
We provide a motivating example, exhibiting similar behavior to all other subjects in our data set, to justify our use of the GMM to standardize RNA-seq data in this work. Figure \ref{fig:Gauss_mix_exp}(a) shows the distribution of $\log_2(\text{FPKM})$ values of gene expression in healthy tissue, provided from GTEx, as a purple curve. It also includes the fitted Gaussian curve  with mean = $8.76$ and standard deviation = $2.08$ in red. However,  the  purple curve is asymmetric and suggests the presence of two subpopulations. Hence we find that a Gaussian mixture model (GMM) of two components is more useful for identifying local features. It should be noted that the Gaussian fitting proposed in \cite{Hart2013} was sufficient in their work, as they incorporated the  ChiP-seq data and in turn estimated an appropriate cutoff point. This allowed them to ignore the left subpopulation. 

As suggested  by the structure of the $\log_2(\text{FPKM})$ curve,  a fitted Gaussian mixture model has two components with means $3.12$ and $8.87$, and standard deviations $2.61$ and $2.11$, respectively (see Figure \ref{fig:Gauss_mix_exp}(b)). Comparing  with Figure \ref{fig:Gauss_mix_exp} (a), we observe that one of the components has very similar parameters to the standard Gaussian fitting.  The presence of the second subpopulation suggests that the incorporation of GMM is necessary to provide a  complete analysis of the data. 
We next estimate standardized scores for GMM, presented in Section \ref{subsec:gm scores}, and their corresponding distributions, to determine the most appropriate score for further analysis.  

\begin{figure}[h!]
				\centering
			
				\includegraphics[width=4.5in,height=3in]{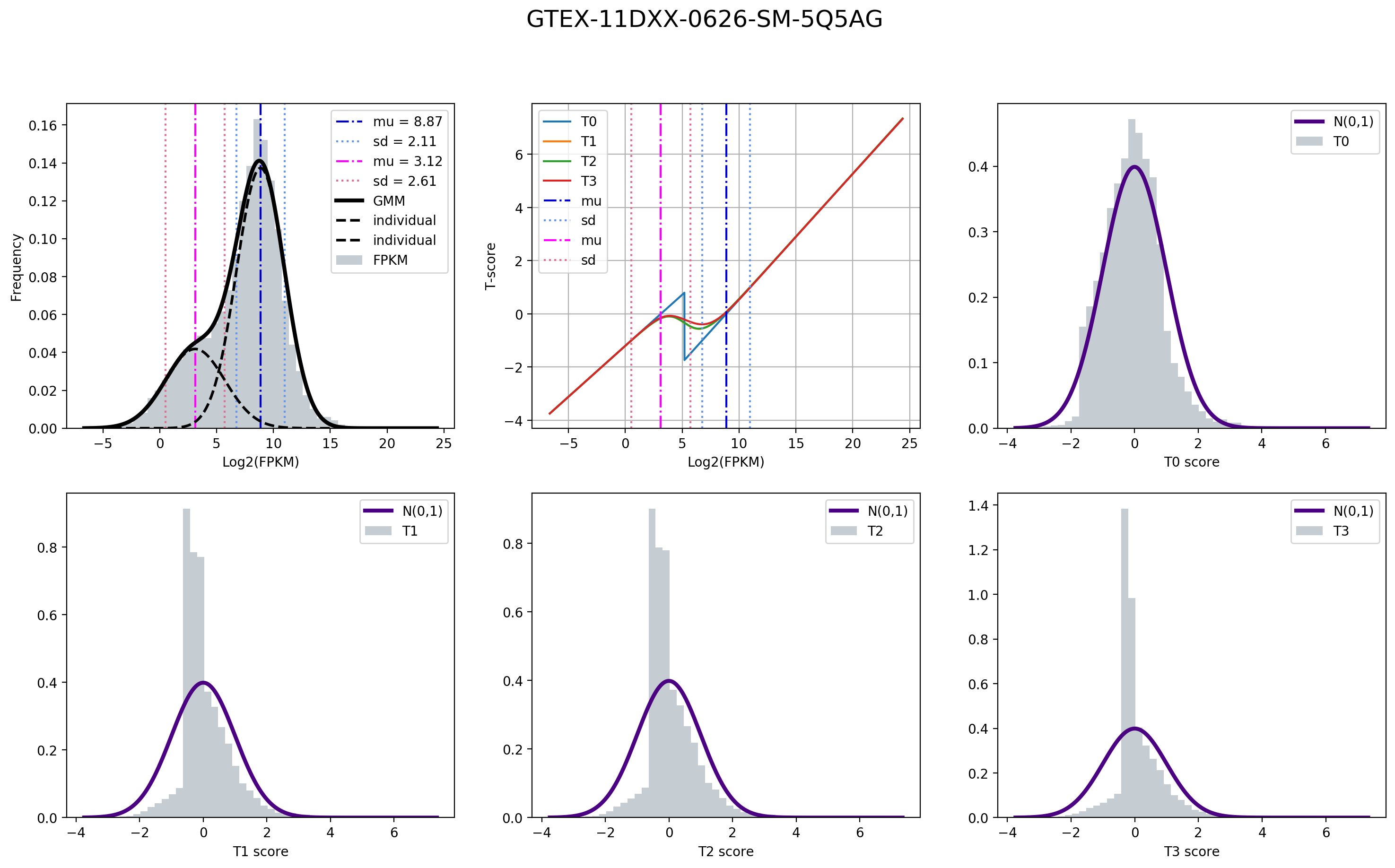} 
				\caption{The histograms and fitted Gaussian curves for all four scores defined in Section \ref{subsec:gm scores}, for a sample subject from the GTEx data set. \label{fig:tscores}}
			\end{figure}

Figure \ref{fig:tscores} shows the histograms of all four scores defined in equations  \eqref{eqn:T0}-\eqref{eqn:T3}, and a fitted Gaussian curve for each, for the sample subject from Figure \ref{fig:Gauss_mix_exp}. It can be observed that the distribution of $T_0$ is symmetric and fits the data well, as we desire from  standardized scores. The readers may expect that gene expressions of different individuals require different types of standardized scores for analysis purposes. However, we have found in all data samples used in this work, that $T_0$ is sufficient for further investigation. The reason  is that  out of these four scores, only the $T_0$ score enjoys a very symmetric normal distribution. Hereafter we restrict our analysis to $T_0$ scores.

\subsection{Mapper algorithm} 
\label{sec:mapper}

To represent our data graphically, we use a topological algorithm, known as Mapper, developed by Singh, Memoli, and Carlssson in \cite{Singh2007}. Mapper, based on a generalized Reeb graph \cite{Reeb1946}, is a tool used for data analysis and visualization, which relies upon a specified filter function to guide the clustering of data points. The filter function assigns a scalar value to each subject, after which subjects are sorted into overlapping bins, according to their corresponding filter function output. Next subjects within each bin are clustered together, according to a specified clustering algorithm, e.g. DBSCAN or agglomerative clustering, to form nodes of the graph. Two nodes are then connected by an edge if they contain one or more subjects in common. 

Thus, Mapper requires the choice of a filter function and clustering algorithm, as well as the specification of input parameters: $b$, the number of filter output bins of equal length; $p$, the percent of overlap between adjacent bins; and $\epsilon$, the scale parameter used in the clustering algorithm, with the metric Euclidean distance. 
We considered several filter functions, including maximum, minimum, and mean correlation, defined for a subject $x$ to be the maximum, minimum, or mean correlation, respectively, between $x$ and all other subjects in the data set. When applying the algorithm to a data set composed of tumor and healthy subjects of a given cancer type, we find that mean correlation is most effective in producing a meaningful graphical structure that largely separates tumor and healthy subjects. Thus, we focus primarily on the mean correlation filter function in this work.

We vary the number of bins, $b$, in the interval $60 \leq b \leq 110$ and vary the parameter $p$ in the interval $30 \leq p \leq 80$. We use the agglomerative clustering algorithm to cluster subjects within each node. The FPKM expression levels and $T_0$ scores are on different orders of magnitude, so for the parameter $\epsilon$ used in clustering data points, we let $600 \leq \epsilon \leq 1000$ when using $T_0$ scores, and let $1\times 10^6 \leq \epsilon \leq 2\times 10^6$ when using FPKM levels for comparison. Since the Mapper algorithm is sensitive to parameter variation, we test the robustness of the graphical structure to variation in parameter values in Section \ref{subsec:sensitivity analysis}.

\subsection{Position Index (PI) \label{subsec:pi}}


In order to classify the position of subjects in a Mapper graph, we define the position index (PI), which assigns a value of $1, -1,$ or $0$ to a subject, according to its position in the Mapper graph. The PI is applicable for graphs that have a strand-like structure, without significant branching, which are the only types of graphs that Mapper produced on the $T_0$ scores and pertinent choices of parameters  of the RNA-seq data used in this study. 

We assign a PI of $+1$ to all subjects that appear in the $13$ nodes on the left side of the visual representation of the Mapper graph. The number $13$ is chosen to account for roughly $\frac{1}{4}$ of the total nodes using our baseline set of parameters. Similarly, we assign a PI of $-1$ to all subjects that appear on the right side of the Mapper graph, and we assign $0$ to all remaining subjects. 

\subsection{Gene Expression Analysis}
\label{sec:gene_analysis}

For the RNAseq transcriptome analysis, variation in gene expression was analyzed using the guideline of DESeq2, an R software package \cite{Love2016}.  The differential gene expression testing method in the package makes use of the negative binomial distribution to model the count data and fit the model by generalized linear models to estimate parameters. The method generates various results to analyze the differential gene expressions and  among them we implement the absolute value of the logarithmic fold changes (LFCs). The fold change between two groups is the ratio of their counts data. For the detection of differential genes DESeq2 tests the null hypothesis 
that the LFC between two groups for a gene’s expression is  zero.


As for Gene Ontology (GO) term analysis, we used the Enrichr platform  \cite{Xie2021, Kuleshov2016, Chen2013} by selecting significantly up or downregulated genes between (1) PI=-1 vs. healthy, and (2) PI=+1 vs. healthy subjects. We selected genes showing the absolute value of the LFCs  $0.90$ or more (i.e. $2^{0.90} = 1.87$ or higher differential expression levels) in the group (1) and (2), yielding 2,885 and 2,923 genes, respectively. Among them, we subcategorized the shared genes between the group (1) and (2) (1,727 genes) and unique genes in (1) (1,158 genes) or (2) (1,183 genes). Each of these three gene lists was loaded at the input data window in the \href{https://maayanlab.cloud/Enrichr/}{Enrichr site},  
and the analyzed results  using the ``GO Biological Process 2021" or ``KEGG 2021 Human", for GO term analysis and pathway analysis, respectively, were downloaded. The top 10 candidates of GO terms and KEGG signaling \href{https://www.genome.jp/kegg/pathway.html}{pathways}  are shown in Figure \ref{fig:genetic}.

\subsection{Graphical Subject Scores (GSS)}
The structure of the Mapper graph depends on the set of parameters chosen by the user in a brute-force setting. Often it is very challenging to measure the robustness of structure to variation in parameters. In this work, we propose a framework to perform a sensitivity analysis of  different parameter choices using \emph{heat kernel signature (HKS)}. We consider Mapper graphical structures as doubly-weighted graphs $G = (V,M,W)$, as defined in Section \ref{subsec:prelm hks}. For a Mapper graph with $n$ nodes, we consider the ratio of healthy and tumor subjects as the weights of vertices $\{m_i\}_{i=1}^n$. The weight of the edge $e_{ij}$ that connects the two vertices $i$ and $j$ is denoted as $W_{ij}$ and estimated as the number of overlapping subjects between  the two nodes $i$ and $j$. The doubly weighted graph proves to be useful for our analysis, as it can incorporate various levels of information of the underlying data set encoded in the Mapper graph. We then compute the graph Laplacian and the corresponding eigenpairs, as defined in Section \ref{subsec:prelm hks}. 
The HKS of each node is computed using Equation \eqref{eq:hks}.  One subject can belong to more than one node, so their GSS is computed as the sum of the HKSs associated with those nodes. The Mapper structure relies on three parameters: the number of filter output bins $b$, the percent of overlap between bins $p$, and the scale parameter for the clustering $\epsilon$. We vary one of them while keeping the other two fixed for the sensitivity analysis.

\section{Results \label{sec:results}}

\subsection{Data Set}
We apply the methodology described in the previous section to RNA-seq data sets, obtained from lung tissue. The data sets contain sequencing data from healthy tissue and from tumor tissue. The healthy data was collected as part of the Genotype Tissue Expression project (GTEx) \cite{Gtex2013,Gtex2015}, and the tumor data was collected from The Cancer Genome Atlas (TCGA) \cite{TCGA}. The data was obtained from the published data in \cite{Wang2018}, so it is processed and normalized according to the process described in \cite{Wang2018}, in order to correct for batch effects and to  allow for comparison between RNA-seq data from two sources. Combining the two data sets, we have 814 total subjects, with 314 healthy subjects from GTEx, and 500 tumor subjects from TCGA. Every subject contains an RNA-seq expression level, reported in FPKM (Fragments Per Kilobase Million), for 19648 distinct genes.

\begin{figure}[h!]
	\centering
	\includegraphics[width=0.45\linewidth]{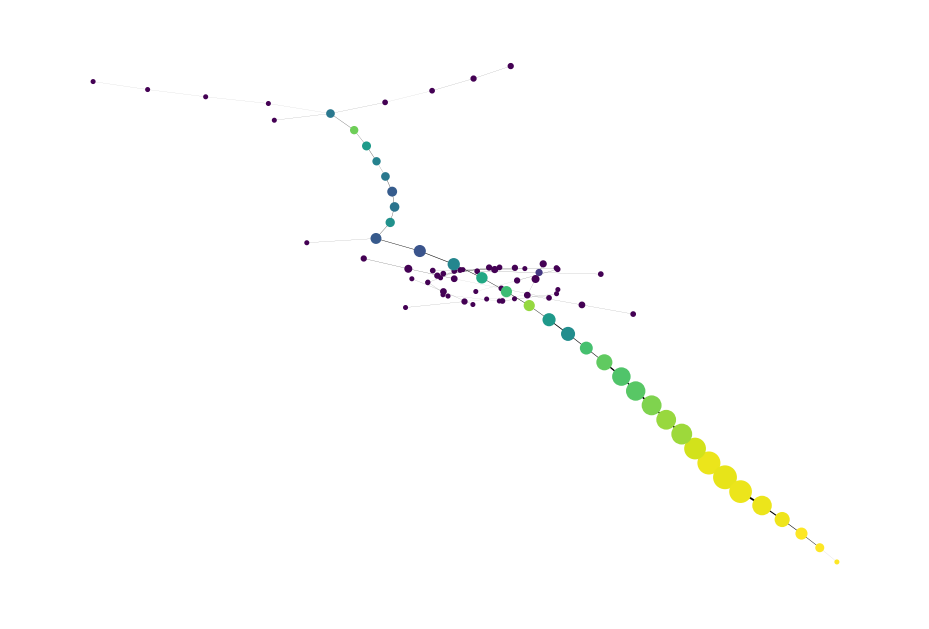} 
	\includegraphics[width=0.45\linewidth]{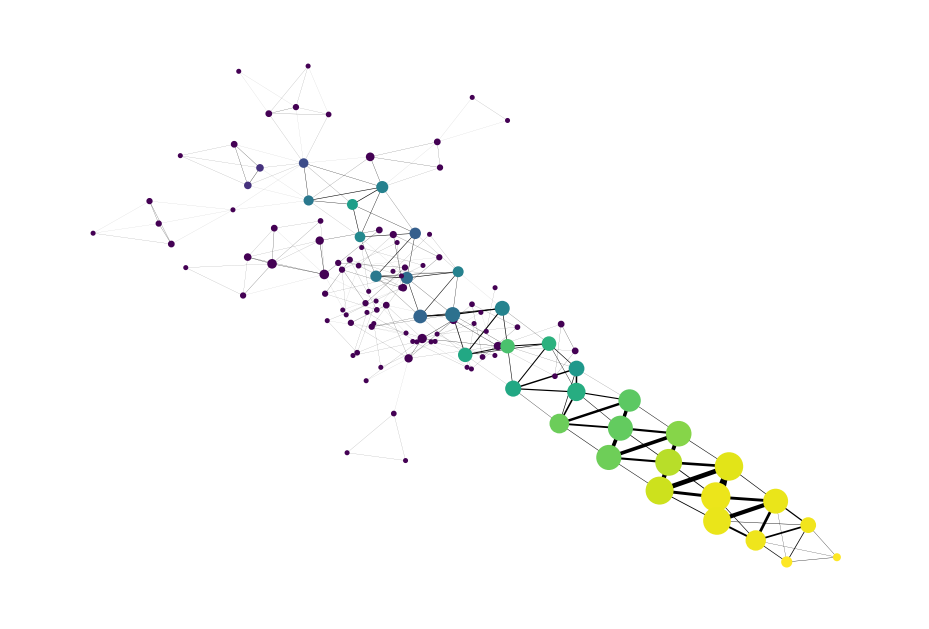} 
	\includegraphics[width=0.45\linewidth]{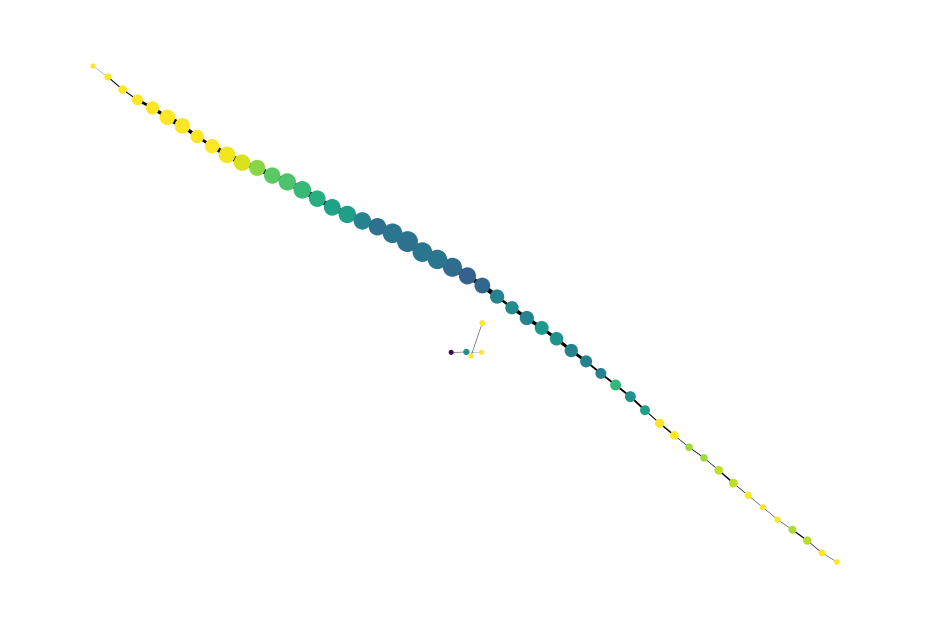} 
	\includegraphics[width=0.45\linewidth]{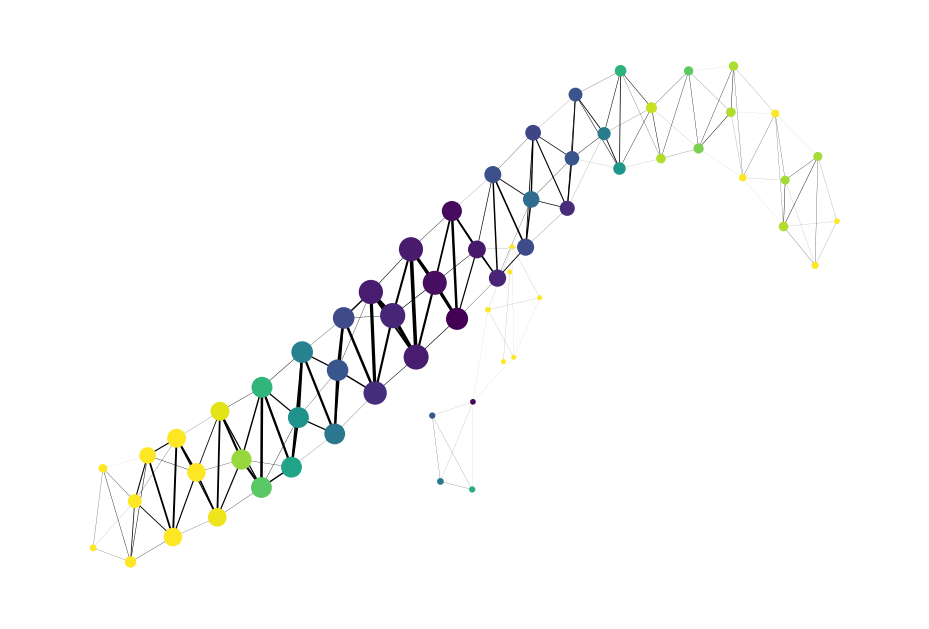} 
	\caption{The top row shows Mapper graphs constructed using FPKM data, and the second row shows Mapper graphs constructed using $T_0$ scores. The FPKM data required a larger $\epsilon$, so we use $\epsilon=2e6$ in the top row, and $\epsilon=600$ in the bottom row. The first column uses the parameter $p=50$, and the second column uses $p=70$. In all cases, we fix $b=80.$ The colors indicate the percentage of tumor subjects versus healthy subjects within each node, with dark purple representing 100\% healthy subjects and bright yellow representing 100\% tumor subjects.}
	\label{fig:mapper_fpkm_t0}
\end{figure}

\subsection{Data Visualization}

Figure \ref{fig:mapper_fpkm_t0} shows a comparison of the Mapper graphs when using FPKM data, displayed in the top row, and when using $T_0$ scores, displayed in the bottom row.  The colors indicate the percentage of tumor subjects versus healthy subjects within each node, with dark purple representing 100\% healthy subjects and bright yellow representing 100\% tumor subjects. Both of the $T_0$ score graphs have a long strand-like appearance, with highly connected nodes in the $p=70$ case, due to the higher percentage of overlap between adjacent bins. In both $T_0$ cases, there are two distinct groups of yellow nodes, indicating that they contain a large number of tumor subjects, separated by all of the healthy subjects, located in the darker nodes near the center of the graphical structure.

Note that when we compare the Mapper graphs using FPKM data versus $T_0$ scores, there is much more consistency in the graphical structure constructed from $T_0$ scores. There is a higher level of connectivity between nodes with $p=70$, but both graphs have a strand-like structure with similar coloring throughout.  Using FPKM data, the graphs for $p=50$ and $p=70$ look particularly different in the regions containing healthy subjects in each case. 
Notice that when using $T_0$ scores the two outer ends of the strand-like graphs are primarily yellow in color, indicating that they contain mostly tumor subjects. The middle of the graph is primarily purple in color, indicating that the centrally located nodes contain mostly healthy subjects. The separated groups of tumor subjects provided the motivation for the position index (PI) described in Section \ref{subsec:pi}, as a method to distinguish between these two groups and for performing further investigations using more traditional differential gene expression analysis.


\subsection{Mapper algorithm} 
\label{sec:mapper}

\begin{figure}[h!]
	\centering
	\includegraphics[width=0.9\linewidth]{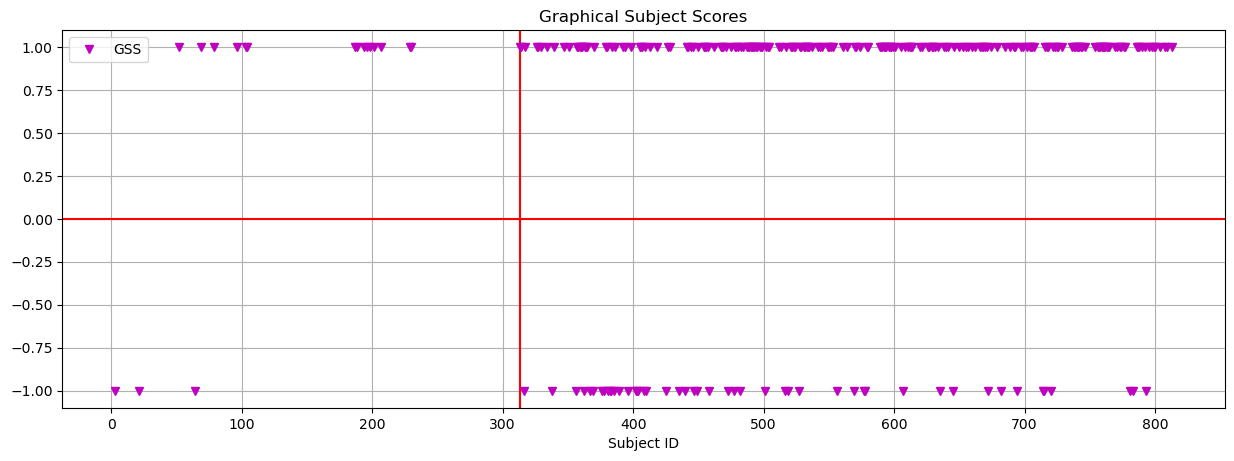} 
	\caption{Position index (PI) of +1 and -1 are shown for healthy subjects, with subject ID smaller than 314 (to the left of the vertical red line), and tumor subjects, with subject ID larger than 314. The PIs are computed from the Mapper graph using $T_0$ scores, with $b=80$, $p=50$ and $\epsilon=600$.}
	\label{fig:GSS_t0}
\end{figure}

 We computed the PI for the subjects clustered in the Mapper graphs using the parameter values $b=80, p \in \{50,70\}$, and $\epsilon\in \{ 600,700 \}$. We found consistency in the PI across these parameter sets.  Thus, using the PI allowed us to distinguish between tumor subjects clustered in the nodes on either end of the Mapper graph.  Figure \ref{fig:GSS_t0} displays the PI for the strand-like Mapper graph that uses $T_0$ scores, with the parameter values $p=50$ and $\epsilon=600$. The vertical red line separates the healthy subjects, on the left, from the tumor subjects, on the right. Note that almost all of the subjects on the two ends of the Mapper graph are tumor subjects, producing the yellow color that we see on the lower left graph in Figure \ref{fig:mapper_fpkm_t0}. In the next section, we describe the results from a genetic analysis that we performed on the two distinct groups of tumor subjects with PI $= +1$ and PI $= -1$, to identify any distinct genetic differences between these two groups. We also explore significant differences between each group of tumor subjects and the healthy subjects (i.e.~those clustered in the center of the Mapper graph). 
 
\begin{figure}[h!]
	\centering
	\includegraphics[width=0.9\linewidth]{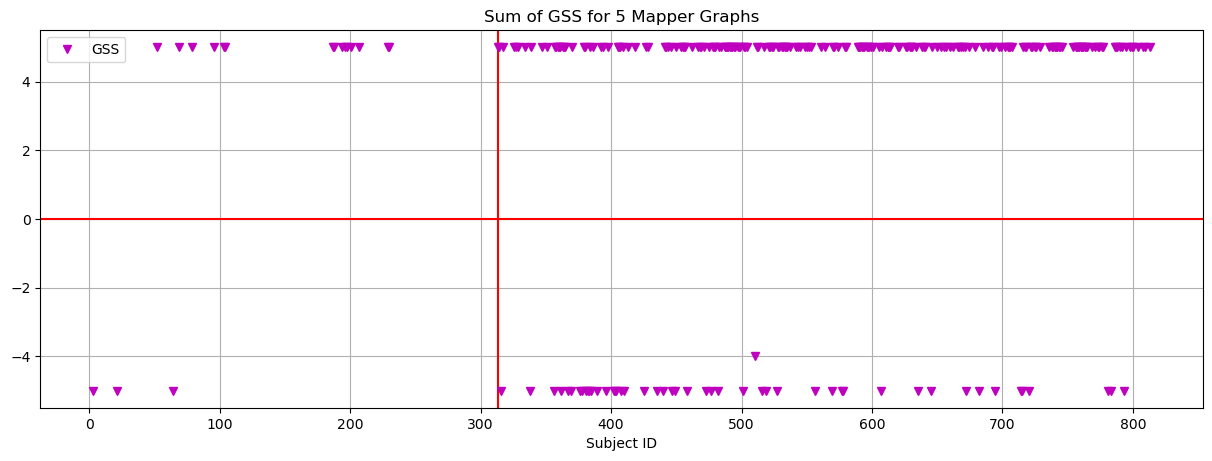} 
	\caption{The sum of the position indices (PIs)  from five different Mapper graphs, with the parameter $\epsilon$ varying in the interval $[600,1000]$. The healthy subject have subject ID smaller than 314 (to the left of the vertical red line), and tumor subjects have subject ID larger than 314. All PIs are computed from the Mapper graphs using $T_0$ scores, with $b=80$ and $p=50$.}
	\label{fig:GSS_t0_sum}
\end{figure}

  We compare the PI for all subjects in Mapper graphs as we vary the parameter $\epsilon$, the scale parameter used in the agglomerative clustering algorithm, in increments of 100 within the interval $\epsilon \in [600,1000]$. Figure \ref{fig:GSS_t0_sum} displays the sum of the PI over the Mapper graphs using these five distinct $\epsilon$ values. We observe that for all but one subject, the sum of the PI is 5, indicating a very high level of consistency in the location of the subjects in the Mapper graph, robust to variation on the local clustering scale within the Mapper algorithm. We perform a more extensive sensitivity analysis of the Mapper structure with a varying set of parameters using HKS in Section \ref{subsec:sensitivity analysis}.  
  
\begin{figure}[htbp]
				\centering
				\includegraphics[width = 1.75in,height=1.75in]{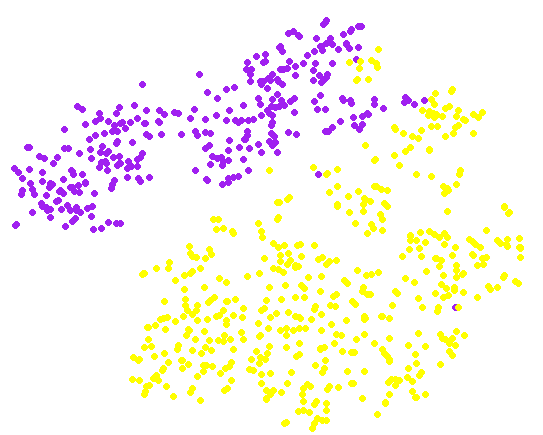}
				\includegraphics[width = 2in,height=1.75in]{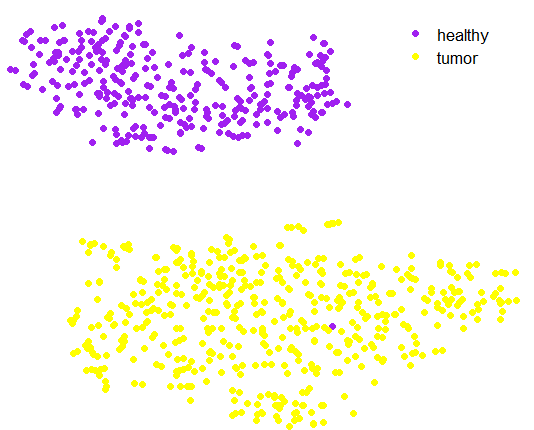} 
				\caption{The t-SNE clusters constructed using lung FPKM data (left) and using lung $T_0$ scores (right).  \label{fig:tsne}}
			\end{figure}

 Finally, we compare Mapper graphs with one of the most commonly used technique to visualize RNA-seq data, t-SNE (t-distributed Stochastic Neighbor Embedding) \cite{Hinton2002, maaten2008}. Although t-SNE is useful in clustering in low-dimensional space, it is unable to produce continuous structures in visualizing gene expression profiles (see \cite{Wang2019} and references therein).  In our analysis, t-SNE can accurately separate healthy and tumor into two different clusters using both FPKM and $T_0$ scores (See Figure \ref{fig:tsne}). However, the two trajectories generated by the PI have not been observed in these clusters.

\subsection{Genetic Analysis}

\begin{figure}[htbp]
	\centering
	\includegraphics[width=0.5\linewidth]{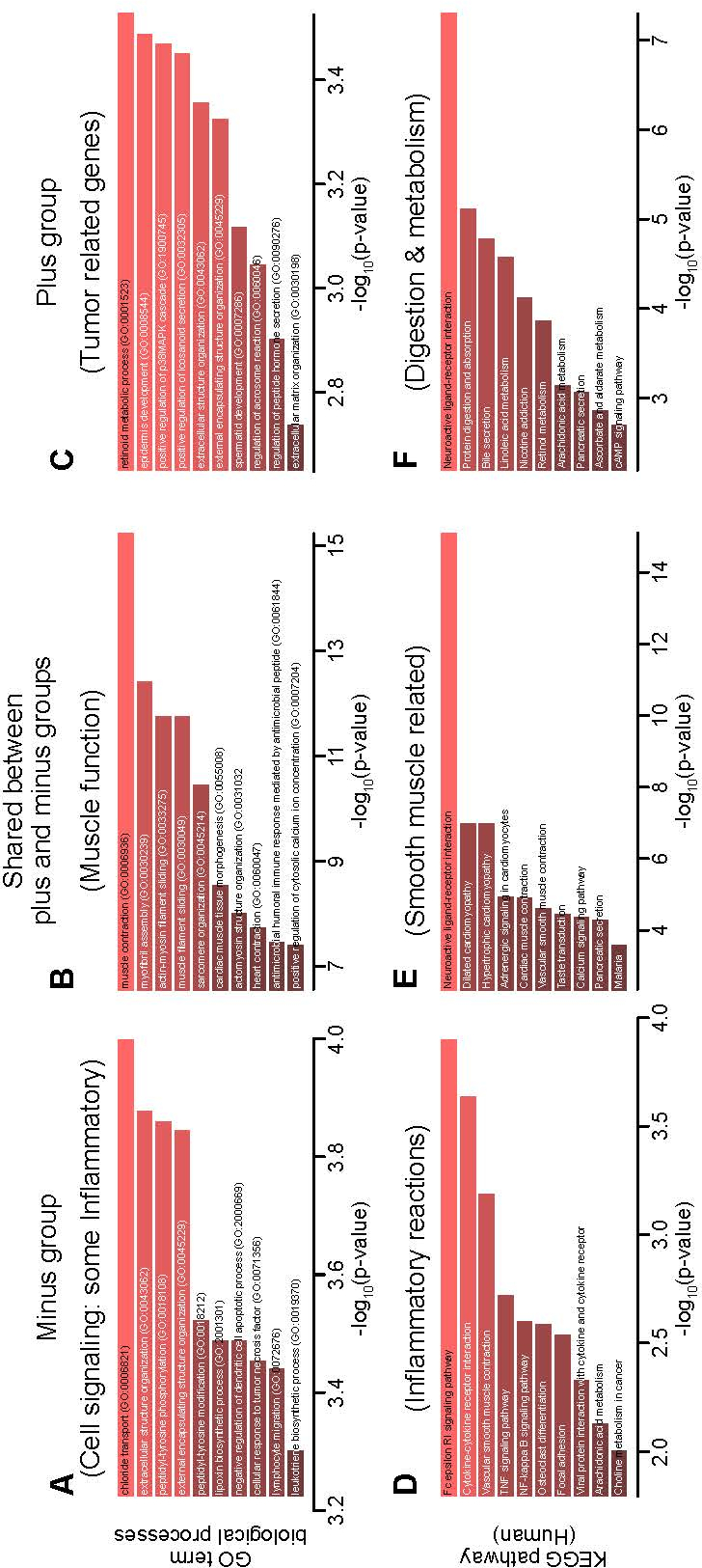} 
	\caption{The biological functions corresponding to the significant genes found uniquely in the PI=-1 group (panels A and D), in the PI=+1 group (C and F), and shared between the +1 and -1 groups (B and E). The top row (A-C) shows the biological processes found using GO term analysis, and the bottom row (D-F) shows the signaling pathway analysis found using KEGG. The x-axes indicate the p-values of significant enrichment. Parentheses indicate the summary of the enriched processes in each panel.}
	\label{fig:genetic}
\end{figure}

We then ran the differential gene expression analyses by using the DESeq2 R package, followed by GO term/KEGG pathway analyses (see Section \ref{sec:gene_analysis}).
As we listed the genes that are significantly differentially expressed between the case and control of lung tumor, we found large overlaps of the genes in the groups of PI = +1 and PI = -1 (GO term analysis and KEGG pathway analysis: Supplementary File 1). Under the GO term analysis, these shared genes revealed enriched clusters with muscle function (e.g., actin family, and myoglobin heavy/light chains), where many genes are upregulated in the cells in both PI = -1 and +1 groups (Figure \ref{fig:genetic}; Supplementary file 2). 

In the genes uniquely grouped in PI = -1, GO-terms are more enriched in cell signaling and inflammatory reaction. KEGG pathways for the PI = -1 genes are also enriched in inflammatory reactions, and many of these genes were upregulated (Figure \ref{fig:genetic}; Supplementary file 1). In contrast, the genes in the PI = +1 group condensed in tumor-related functions (retinoid metabolism, epidermis development, p38MAPK enhancement), and KEGG showed retinol metabolism and some diet digestion and metabolisms (Figure \ref{fig:genetic}; Supplementary files 1 and 2). We discuss the utility of the  information and the hidden biological significance revealed by the Mapper graph.

The Mapper graph indicated that the tumor subjects’ cells were clustered in two groups on either end of the strand-like graphical structure, which we labeled with  PI = +1 and -1, while the cells from healthy individuals were largely clustered in the center. The two PI groups suggest two possible processes for forming the lung cancer. For example, the PI = -1 group is strongly clustered with the tumor cells (majorities of nodes are yellow colored; the bottom row, left panel in Figure 4), suggesting that the cells upregulating the inflammatory reactions are primarily cancer cells (Supplementary files 1 and 2). In contrast, the PI= +1 group consists of mixture of healthy and cancer cells (green and yellow nodes; the bottom row, left panel in Figure 4), indicating the cells in this PI= +1 arm trajectory, although still ‘high risk,’ are not as high as the PI= -1 arm pathway for the lung cancer. Uniquely diversified gene expressions in PI= +1 group are enriched with signaling pathways associated with tumors and nicotine consumption (Figure 5; Supplementary file 2), suggesting possible environmental factors (such as smoking) and tumor gene interactions.

The two trajectories represented by PI= -1 and +1 have not been suggested in lung cancer studies thus far, which are based on t-distributed stochastic neighbor embedding (t-SNE) \cite{Hinton2002,maaten2008},  or other clustering methods. Our pipeline presents a new path to reveal individualized treatment strategies for lung cancer as well as, more generally, to interpret bulk RNA-seq based studies.
The vertices in PI= -1 and +1 are continuously connected by edges; therefore, we can predict the risk state of ‘healthy cells’: if the cells belong to the vertices close to PI= -1 or +1 group, these cells are at higher risk to be cancer cells than the cells in the middle of the healthy cell cluster. In a biological context, if the expressions of muscle-related genes plus inflammatory-related genes are upregulated, these cells are at the highest risk. This type of interpretation using vertices-edge patterns is very difficult to obtain using the current popular clustering algorithms, including t-SNE \cite{Hinton2002, maaten2008}.
We also considered shared differentially expressed genes in both the PI= -1 and +1 groups, and we found they are enriched in muscle physiology/differentiation (Figure 5). This result was surprising to us because, as far as we know, the association between muscle contractions and lung cancer has not been well-characterized. One possibility is that actin-myosin and related genes are necessary to be recruited during metastasis. Further investigation of muscle genes in the context of lung cancer may shed light on previously unknown aspects of lung cancer development.

\subsection{Sensitivity Analysis \label{subsec:sensitivity analysis}}

In order to showcase the implication of GSS, we perform a simple sensitivity analysis under different parameter sets that are used to generate Mapper graphs. As discussed in Section \ref{sec:mapper} the Mapper pipeline we consider here typically relies on three parameters, namely the number of  bins  $b$, the percent of overlap $p$, and the scale parameter used in the clustering algorithm $\epsilon$. We estimated the GSS for the subjects in the underlying data set using all possible combinations of the parameter values $b \in \{60, 70, 80, 90, 100, 110\}$,  $p \in \{30, 40, 50, 60, 70, 80\}$, and $\epsilon \in \{600, 700, 800, 900, 1000\}$. 
\begin{figure}[h!]
	\centering
		\includegraphics[width=1\linewidth]{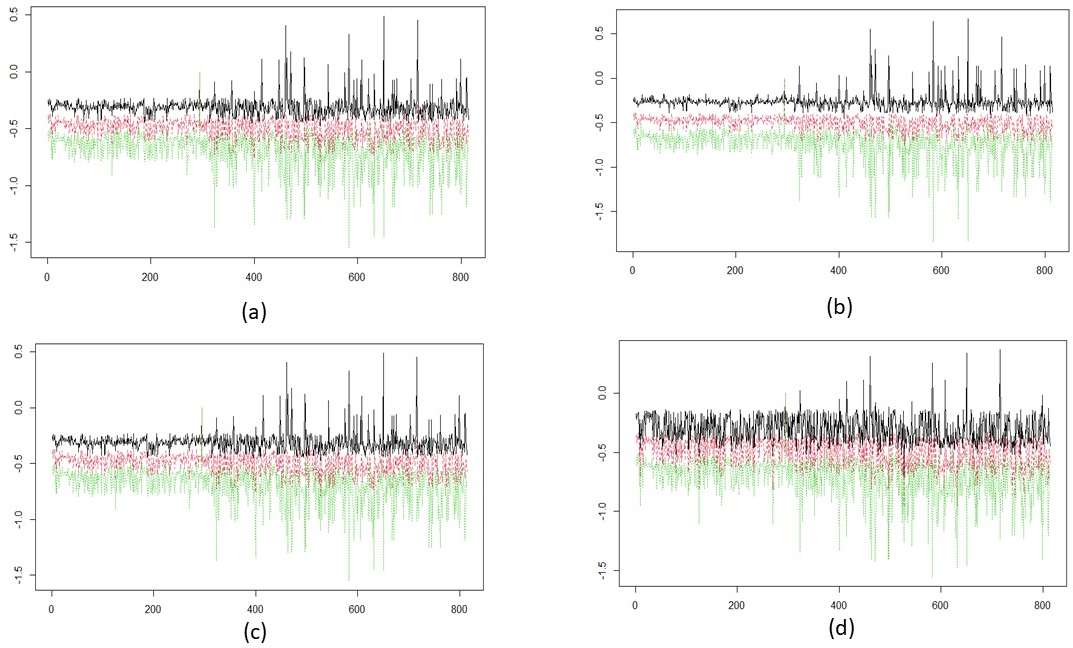} 
	\caption{$95\%$ confidence intervals (CIs) of graphical subject scores (GSSs) are shown for 814 subjects with (a) $\epsilon = 700$ and  $p = 80$, (b) $\epsilon = 800$ and  $p = 80$, (c) $\epsilon = 900$ and  $p = 80$, and (d) $\epsilon = 1000$ and  $p = 80$. The CIs were estimated by considering  all possible values of the parameter $b$. The horizontal axis represents  the subject index and vertical axis represents the GSSs. The red line is for the mean of the observations, black line is for the upper confidence limit, and green one is lower confidence limit. The first 314 subjects are healthy subjects, and we observe a distinguishable pattern in their GSSs with respect to the other 500 subjects. }
	\label{fig:epsilon_g}
\end{figure}

In order to perform the sensitivity analysis we  vary one parameter while leaving the other two fixed. We then estimate the $95\%$ confidence interval from the empirical distribution of the GSSs of each subject in the data set. For the sake of brevity, we present some selected results that show robustness of the parameters.
We found consistency in the  GSSs and, consequently, tighter confidence intervals across the parameter sets $b = \{60, 70, 80, 110\}$ and  $p = \{40,50,60,70\}$. We observe that several combinations of these parameter sets produce Mapper graphs that are very robust to changes in $\epsilon$. Furthermore, the GSSs show distinguishable pattern between healthy and tumor subjects. Precisely,  we observe less variation in the scores of healthy subjects and more in that of tumor subjects. This pattern was anticipated, as the tumor subjects were clustered in the nodes on either end of the Mapper graph, and healthy subjects are located at the center of the graph.



\begin{figure}[h!]
	\centering
	\includegraphics[width=1\linewidth]{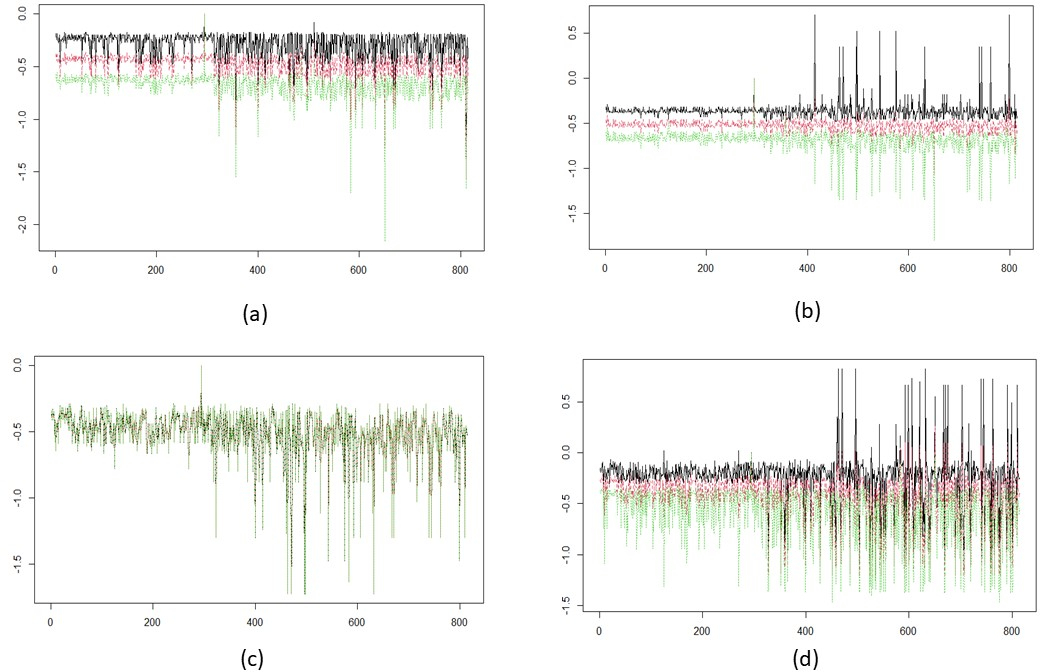} 
	\caption{$95\%$ confidence intervals (CIs) of graphical subject scores (GSSs) are shown  with (a) $b = 60$ and  $p = 80$, (b) $b = 70$ and  $p = 80$, (c) $b = 80$ and  $p = 80$, and (d) $b = 110$ and  $p = 80$. The CIs were estimated by consider  the parameter $\epsilon \in \{600, 700, 800, 900, 1000\}$. The horizontal axis represents  the subject index and vertical axis represents the GSSs. The red curve is for the mean of the observations, black curve is for the upper confidence limit, and green curve is lower confidence limit. The first 314 subjects are healthy subjects and we observe distinguishable pattern in their GSSs with that of other 500 subjects.}
	\label{fig:r_g_80}
\end{figure}

Figure \ref{fig:epsilon_g} displays the GSSs and corresponding CIs for the parameter values $\epsilon \in \{700, 800, 900, 1000\}$ and $p=80$. For all of the cases in Figure \ref{fig:epsilon_g} we vary the values of $b$ in $\{60, 70, 80, 90, 100, 110\}$  The horizontal axis represents  the subject index and vertical axis represents the GSSs. The red curve is for the mean of the observations, the black curve is for the upper confidence limit, and the green curve represents the lower confidence limit. We observe a tighter confidence range for the healthy subjects' scores, especially in Figure \ref{fig:epsilon_g}(a), (b), and (c). A thicker confidence range is generated in Figure \ref{fig:epsilon_g}(d). For all of these cases we found that the CI is wider for some of the tumor subjects. This is because  the tumor subjects were clustered in the nodes on either end, and HKS values reflect where the subject is clustered in the Mapper graph.  We note also that varying the number of filter output bins, $b$, changes the number of nodes in the Mapper graph, which impacts the GSS for many subjects. 

\begin{figure}[h!]
	\centering
	\includegraphics[width=1\linewidth]{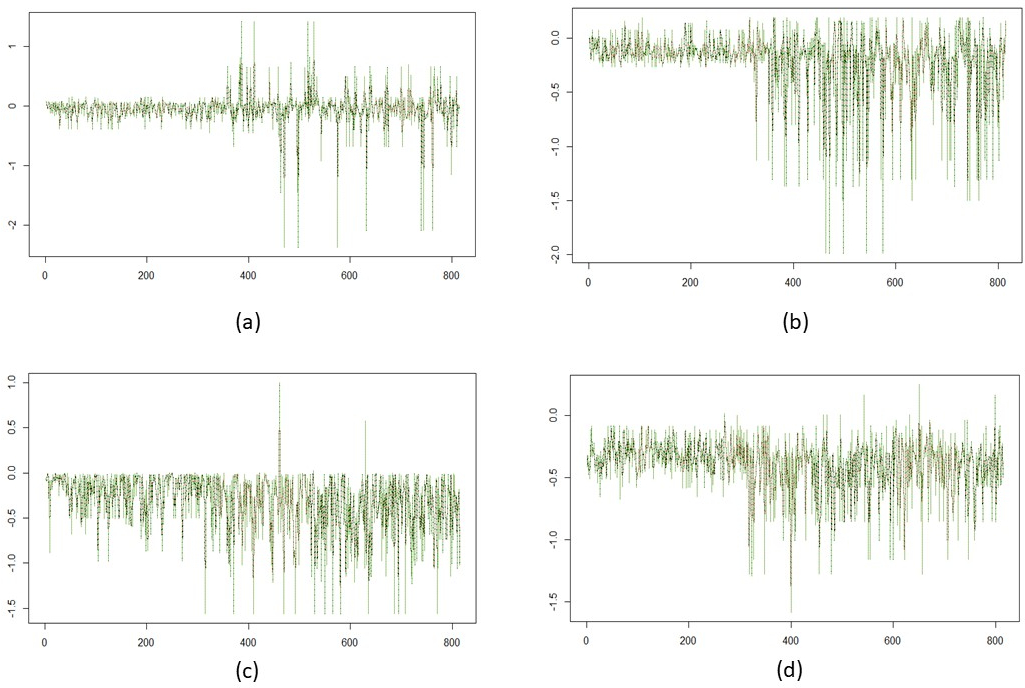} 
	\caption{$95\%$ confidence intervals (CIs) of graphical subject scores (GSSs) are shown  with (a) $b = 80$ and  $p = 40$, (b) $b = 80$ and  $p = 50$, (c) $b = 80$ and  $p = 60$, and (d) $b = 80$ and  $p = 70$. The CIs were estimated by considering  the parameter $\epsilon \in \{600, 700, 800, 900, 1000\}$. The horizontal axis represents  the subject index and vertical axis represents the GSSs. The red curve is for the mean of the observations, black curve is for the upper confidence limit, and the green curve shows the lower confidence limit. }
	\label{fig:r80_g}
\end{figure}

Figures \ref{fig:r_g_80}  and \ref{fig:r80_g} display the GSSs and corresponding CIs for combinations of parameter values $b \in \{60, 70, 80, 110\}$ and $p \in \{ 40, 50, 60, 70\}$. For all of the cases in Figure \ref{fig:epsilon_g} we vary the values of $\epsilon$ in $\{600, 700, 800, 900, 1000\}$.   The horizontal axis represents  the subject index and vertical axis represents the GSSs. The red curve displays the mean of the observations, the black curve is for the upper confidence limit, and the green curve shows the lower confidence limit. Similarly to the results above, in Figure \ref{fig:r_g_80}, we observe a tighter confidence range for the healthy subjects' scores and a wider confidence range for tumor subjects, due to the clustering of  the tumor subjects in the nodes on either end of the Mapper graph.  We note that in Figure \ref{fig:r80_g}, as we vary $\epsilon$, the Mapper graphs are nearly identical, which is why the green, red, and black curves appear nearly overlapping. This demonstrates that the Mapper graphs and the clustering of the subjects are very robust to changes in $\epsilon$ in the parameter regime tested here.

\section{Conclusion and Discussion \label{sec:discussion}}

In this work, we developed a novel framework to analyze RNA-sequencing data using the topological algorithm known as Mapper and a corresponding scoring method, known as the Heat Kernel Signature (HKS). In order to produce a Mapper graph with an informative structure, we first describe the RNA-seq data using a Gaussian Mixture Model (GMM) and corresponding normalization scores. We tested this framework on a data set containing genetic expression levels in lung tissue from tumor and healthy subjects, which revealed a graphical structure with two distinct groups of tumor subjects. The distinct groups are not found using traditional statistical clustering methods, such as t-SNE. 

Subsequent genetic analysis, using the differential gene expression analyses with DESeq2, followed by Enrichr, of the two tumor subgroups and the healthy subjects, identified the most significant genes within each group. This analysis revealed two distinct pathways for the development of lung cancer. One pathway involves retinoid metabolism and digestion, and the other primarily involves inflammatory reactions. The mixture of healthy subjects with the PI=+1 group of tumor subjects suggests that the upregulation of metabolism and digestion-related genes is not as high risk as the upregulation of inflammatory-related genes. Additionally, muscle-related genes were upregulated in both tumor groups, so cells with high expressions of muscle-related genes plus inflammatory-related genes could be at the highest risk. However, without further investigation, we cannot interpret well why muscle signaling is involved in tumor formation. We suspect the actomyosin complex (a part of the muscle signaling) could contribute to metastasis in lung cancer. The separation of the two tumor groups in the Mapper graph allowed us to discover these important genetic distinctions between the groups, which have not been well-characterized previously, to the best of our knowledge. These results suggest a direction for further biological investigation, which could reveal previously unknown lung cancer biomarkers and new mechanisms for the development of lung cancer.

Additionally, we showcased how the HKS can be applied to Mapper graphs in order to provide an associated score for each subject. We call this  signature the graphical subject score (GSS), and it can be used to perform a parameter sensitivity analysis for a given data set, as we have demonstrated in this work. In our data set, the tumor subjects were more sensitive to parameter variation, due to the changing number of nodes as $b$ and $p$ varied, and the clustering of tumor subjects on both ends of the graph. However, the positioning of all subjects was very robust to changes in the clustering parameter $\epsilon$. The GSS is useful for identifying Mapper parameter ranges that produce consistent graphical structures. Since it associates a score for each subject over parameter ranges, this method can be used in future work to allow for more robust statistical inference involving Mapper graphs.

\section*{Acknowledgements}
The work has been partially supported by the NIH COBRE P20GM125508 (FN and MY). This research began at the ICERM Workshop: Applied Mathematical Modeling with Topological Techniques, and we thank Vladislav Bukshtynov, Steven Ellis, Elin Farnell, Hwayeon Ryu, and Sarah Tymochko  for their ideas during early conversations related to this work.

\bibliographystyle{unsrt}      
\bibliography{main}   

\end{document}